\documentclass[prl,reprint,showpacs,showkeys,showpacs,times,superscriptaddress]{revtex4-1}
\usepackage{amsmath,amssymb,amsthm,times,graphics,graphicx,bm,subfigure}

\usepackage[colorlinks,linkcolor=red,citecolor=blue,urlcolor=blue]{hyperref}
\usepackage{enumerate}

\newcommand{\be}{\begin{equation}}
\newcommand{\ee}{\end{equation}}
\newcommand{\ben}{\begin{eqnarray}}
\newcommand{\een}{\end{eqnarray}}
\newcommand{\bes}{\begin{subequations}}
\newcommand{\ees}{\end{subequations}}
\newcommand{\bF}{\begin{figure}}
\newcommand{\eF}{\end{figure}}

\def\ket#1{ | #1 \rangle}
\def\bra#1{{\langle #1 |  }}

\def\tr#1{{\rm{Tr}}\left[#1\right]}

\def\pd2v#1#2#3{\frac{\partial^2 #1}{\partial #2 \partial #3}}

\def \2x2mat#1#2#3#4{
\left( \begin{array}{cc}
#1 &  #2 \\  #3 &  #4
\end{array} \right)
}

\def \i{i}

\begin{document}
\title{Entangling measurements for multiparameter estimation with two qubits}
\author{Emanuele Roccia}
\affiliation{Dipartimento di Scienze, Universit\`a degli Studi Roma Tre, Via della Vasca Navale 84, 00146, Rome, Italy}
\author{Ilaria Gianani}
\affiliation{Dipartimento di Scienze, Universit\`a degli Studi Roma Tre, Via della Vasca Navale 84, 00146, Rome, Italy}
\author{Luca Mancino}
\affiliation{Dipartimento di Scienze, Universit\`a degli Studi Roma Tre, Via della Vasca Navale 84, 00146, Rome, Italy}
\author{Marco Sbroscia}
\affiliation{Dipartimento di Scienze, Universit\`a degli Studi Roma Tre, Via della Vasca Navale 84, 00146, Rome, Italy}
\author{Fabrizia Somma}
\affiliation{Dipartimento di Scienze, Universit\`a degli Studi Roma Tre, Via della Vasca Navale 84, 00146, Rome, Italy}
\author{Marco G. Genoni}
\affiliation{Quantum Technology Lab, Dipartimento di Fisica, Universit\`a degli Studi di Milano, 20133, Milan, Italy}
\author{Marco Barbieri}
\affiliation{Dipartimento di Scienze, Universit\`a degli Studi Roma Tre, Via della Vasca Navale 84, 00146, Rome, Italy}

\begin{abstract}
Careful tailoring the quantum state of probes offers the capability of investigating matter at unprecedented precisions. Rarely, however, the interaction with the sample is fully encompassed by a single parameter, and the information contained in the probe needs to be partitioned on multiple parameters. There exist then practical bounds on the ultimate joint-estimation precision set by the unavailability of a single optimal measurement for all parameters. Here we discuss how these considerations are modified for two-level quantum probes - qubits - by the use of two copies and entangling measurements. We find that the joint estimation of phase and phase diffusion benefits from such collective measurement, while for multiple phases, no enhancement can be observed. We demonstrate this in a proof-of-principle photonics setup.
\end{abstract}

\maketitle

\section{Introduction}
Monitoring a system, being it for fundamental studies or for sensing applications, always requires to determine a set of physically meaningful parameters, that summarise the essentials of its behaviour. 
As the evolution of such a system influences all these quantities at once, their simultaneous estimation is highly desirable for tracking changes in time. There exists a fundamental limitation in the fact that these parameters might be associated to conjugated variables in quantum mechanics~\cite{Giovannetti06}. Quantum metrology, once declined in a multiparameter framework, aims at understanding and reaching these ultimate limits~\cite{Paris08,Szczykulska16,Ragy17}.

The evolution parameters find a description within different categories: either unitary or dissipative transformations. Since the effect of a unitary evolution almost always shows up as the introduction of a phase, phase estimation has long represented the core business of quantum metrology~\cite{Holland93,Braunstein94,Mitchell04,Walther04,Giovannetti04,Higgins07,Nagata07,Afek10}; more general instances might demand a description in term of multiple phases or generic parameters characterizing unitary operations~\cite{Dariano97,Ballester04,Ballester04a,Spagnolo12,Humphreys13,Genoni13,Vaneph13,Baum16,Gagatsos16}. Accompanying dissipative phenomena have often being treated as a limiting factor spoiling quantum enhancement~\cite{Dorner09,Kacprowicz10,Genoni11,Datta11,Knysh11,Escher11,Genoni12}, however, there exist cases in which dissipation can provide insight on the system~\cite{Monras07,Adesso09,Chiuri11,Crowley14,Pirandola16}, as it is the case for decoherence microscopy~\cite{Cole09} and thermometry~\cite{Kukso13,Toyli13}. 

Concerning the estimation precision of multiple parameters, the main result in quantum metrology is a bound holding for the covariance matrix for any possible measurement, the so-called Quantum Cram\'er-Rao (QCR) bound~\cite{Paris08}. In its generality, the QCR bound might fail at shedding light on trade-offs arising in the optimal precision for individual parameters in a simultaneous strategy. This observation applies regardless the nature of the parameters~\cite{Gill00}, and, lacking a comprehensive theory, case studies are particularly informative. Attention has been devoted to the simple two-level quantum bit (qubit) instance, which effectively describes many relevant cases~\cite{Brivio10,Vidrighin14}. The estimation of multiple parameters with single copies has been investigated in Refs.~\cite{Gill00,Vidrighin14,Altorio15,Szczykulska17}, that have shown how the information in the probe state is distributed on the different parameters, hence the individual precisions are affected; in a brief summary, the better we estimate one parameter, the worse we are bound to get for the others. 

\begin{figure}[b!]
\includegraphics[width=\columnwidth]{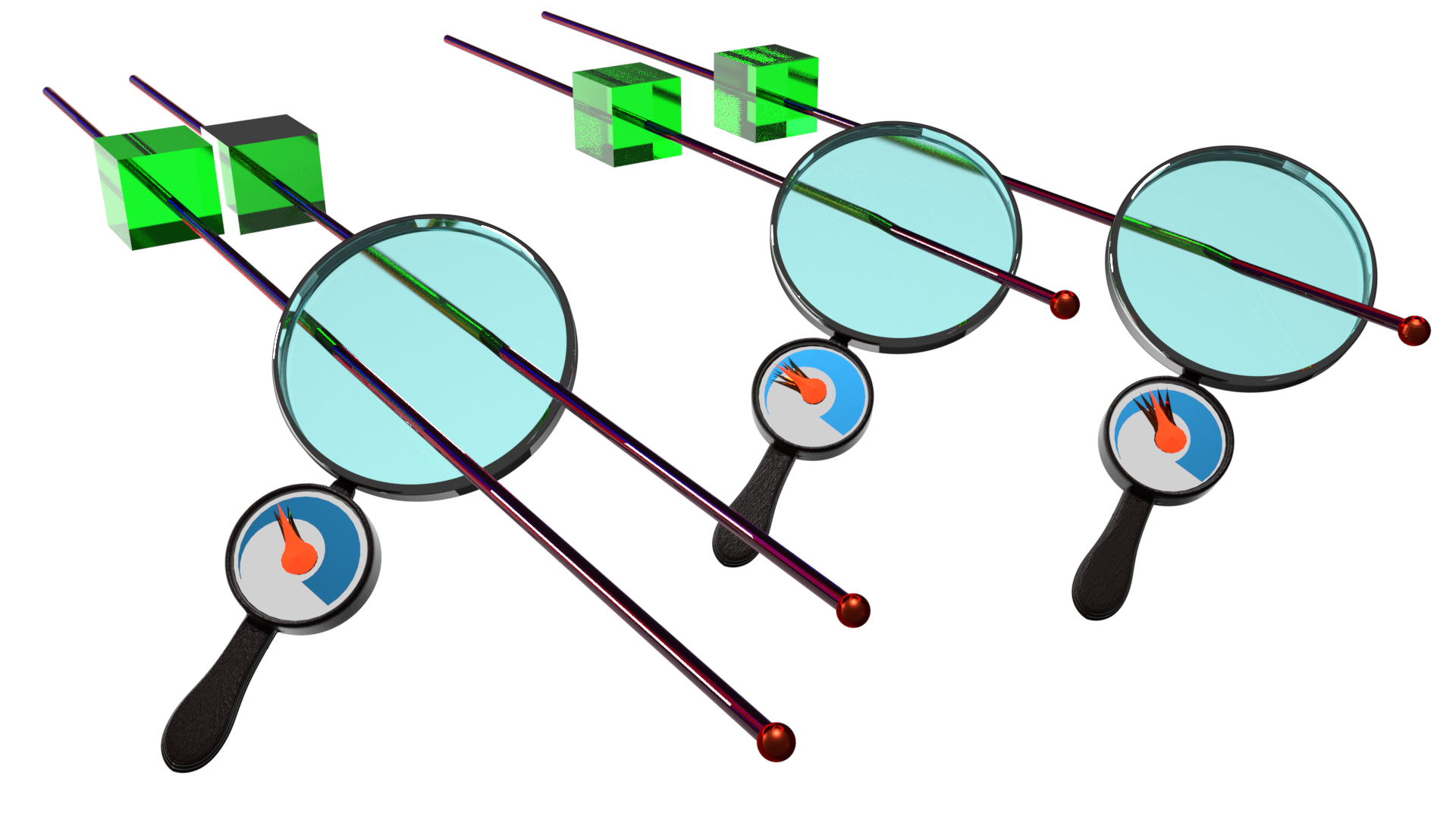}
\caption{Multiparameter estimation with multiple copies. A material is investigated with either a collective or and individual measurement on two probes; the relevant aspects of the sample are codified as parameters to be estimated. Under specific circumstances, the collective strategy may result in a better precision in the simultaneous estimation of the parameters.}
\label{Figura1}
\end{figure}

When dealing with a phase and an associated phase diffusion with qubits, it has been predicted that the use of collective, entangling measurements on two copies at once can mitigate this trade-off in the precisions~\cite{Vidrighin14}. This highlights an important difference with respect to the single-parameter case where entangling measurements are not expected to deliver any advantage~\cite{Giovannetti06}. Conversely, in the joint measurement of phase and phase diffusion, collective measurements can impose less severe compromises, although it should be emphasised that the scaling of the precision with the number of copies is not modified. This is a complementary strategy with respect to preparing the probe state outside the qubit space~\cite{Szczykulska17}.

Here we present the experimental characterisation of the metrology capabilities of a linear-optical entangling gate~\cite{Langford05,Schmid09} for multiparameter estimation at the proof-of-principle level. We apply detector tomography to our device~\cite{Lundeen09} to quantify how the device implements the estimation of (i) phase and phase diffusion parameters, and (ii) multiple phases using two qubits at the time. We find that the advantage offered for (i) is highly sensitive to imperfections, and that, more importantly, it is not of general nature, since it does not hold for (ii).  Our study is both a critical assessment of the cost/benefit balance of adopting collective strategies, and a further demonstration that metrologic trade-offs are better understood in terms of the evolution uphill rather than the final measurement downhill~\cite{Crowley14,Durkin}.

\section{Results}
{\it Multiparameter quantum metrology.} We are concerned with the problem of comparing the two measurement strategies sketched in Fig.~\ref{Figura1}. We can use two probes at the time, and perform a collective measurement, or use single probes and individual measurements.  

The generic estimation problem starts with the interaction of the probe in the initial quantum state $\ket{\psi_0}$, which is transformed by the evolution inside the sample in the state $\rho_{\vec\phi}$, where $\vec \phi$ identifies a set of $n$ parameters characterising the evolution, might contain both unitary and dissipative parameters. Any measurement has outcomes $s$ each occurring with a probability $p(s|\vec\phi)$, which deliver quantitative information on the parameters in the form of a vector of estimators $\vec\phi'$; we will focus, as customary, on the case of unbiased estimators where the expectation value of the estimators coincide with the actual values: $\langle\vec\phi'\rangle=\vec\phi$, and on the problem of local estimation concerning an improvement over an initial estimate. Peaked distributions are expected to be more informative, as small shifts then signal small variations in the parameters; also, correlations may appear in the estimators due to the form of the probabilities. These observations can be made quantitative by looking at the Fisher Information (FI) matrix: $F_{ij}(\vec \phi)=\sum_s \left(\partial_{\phi_i}p(s|\vec\phi)\partial_{\phi_j}p(s|\vec\phi)\right)/p(s|\vec\phi)$, which bounds the covariance matrix $\Sigma _{ij} ={\langle(\phi'_i-\phi_i)(\phi'_j-\phi_j)\rangle}$ through the classical Cram\'er-Rao bound $\Sigma\geq F^{-1}(\vec \phi)/M$ ($M$ being the overall number of measurements performed) in the asymptotic regime of a large number of repetitions. For any given parameter $\phi_j$, the diagonal value $F^{-1}_{jj}(\vec \phi)$ set a limit to the associated variance that encompasses both the dependence of $p(s|\vec \phi)$ on $\phi_j$ and its correlations with the remaining parameters; this represents an effective FI as $F^{\rm eff}_{jj}(\vec\phi)=1/F^{-1}_{jj}(\vec \phi)$.   

There exists a fundamental limit to the achievable classical FI, based solely on how the output state depends on the parameters. Given the Symmetric Logarithmic Derivative (SLD) operators $\{L_i\}$ defined as: $2\partial_{\phi_i}\rho_{\vec\phi}=L_\i\rho_{\vec\phi}+\rho_{\vec\phi}L_i$, these impose the QCR bound: $F{\leq}H$, where the quantum FI matrix is now defined as $H_{ij}{=}\tr{\rho_{\vec\phi} \{ L_i, L_j\}/2}$, with $\{A,B\}$ denoting the anti-commutator. While for the single-parameter case there is always the guarantee of reaching the optimal FI by choosing a measurement along the eigenbasis of $L_1$, this feature is lost for multi-parameter problems due to the possible non-commutativity of the different SLDs~\cite{Ragy17,Gill11a}. In fact, even if the SLD operators do not commute, this does not directly imply the impossibility of achieving the QCRB for all the parameters characterizing the quantum state. However the following weaker condition
\begin{align}
\tr{ \rho_{\vec\phi} [ L_i , L_j ] } = 0 \label{eq:cond},
\end{align}
has been proved to be necessary and sufficient \cite{Matsumoto02,Ragy17}. In plain words the multi-parameter QCR bound can be achieved if and only if the expectation value of the commutator of the SLD operators on the probe state is equal to zero.

One could then ask which are the ultimate performances in multi-parameter estimation when a fixed measurement strategy is considered. It has been suggested that insight on such problems could come from considering the quantity 
\be 
\kappa=\sum_j \frac{F^{\rm eff}_{jj}/m}{\,H_{jj}} 
\label{kappa}
\ee
as a figure of merit~\cite{Ballester04,Ballester04a,Gill00,Vidrighin14}, where $m$ denotes the number of copies of the probe states that are jointly measured. Whenever the condition (\ref{eq:cond}) is satisfied, the optimal measurement, that in general is entangling and acts jointly on an asymptotically large number of copies $m$ of the probe state~\cite{Gill11a}, would saturate the condition $\kappa{\leq}n$ implied by the QCR bound. In the following we will discuss the ultimate limits on $\kappa$ for specific estimation problems and experimentally realisable measurement strategies.

{\it Qubit metrology}. We focus our attention to the qubit case, which is not only relevant for Ramsey interferometry, but for an effective description of $N00N$ states and coherent states in optical interferometry as well~\cite{Vidrighin14,Szczykulska17}. With individual probes at out disposal, the condition $\kappa \leq 1$ holds for any two parameters~\cite{Gill00}, thus failing to meet the optimal condition $\kappa \leq 2$. This has been explicitly shown in experiments by inspecting polarisation detectors for the cases of phase-dephasing~\cite{Vidrighin14}, and two-phase estimation problems~\cite{Altorio16}. 

\begin{figure*}[t!]
\includegraphics[width=\textwidth]{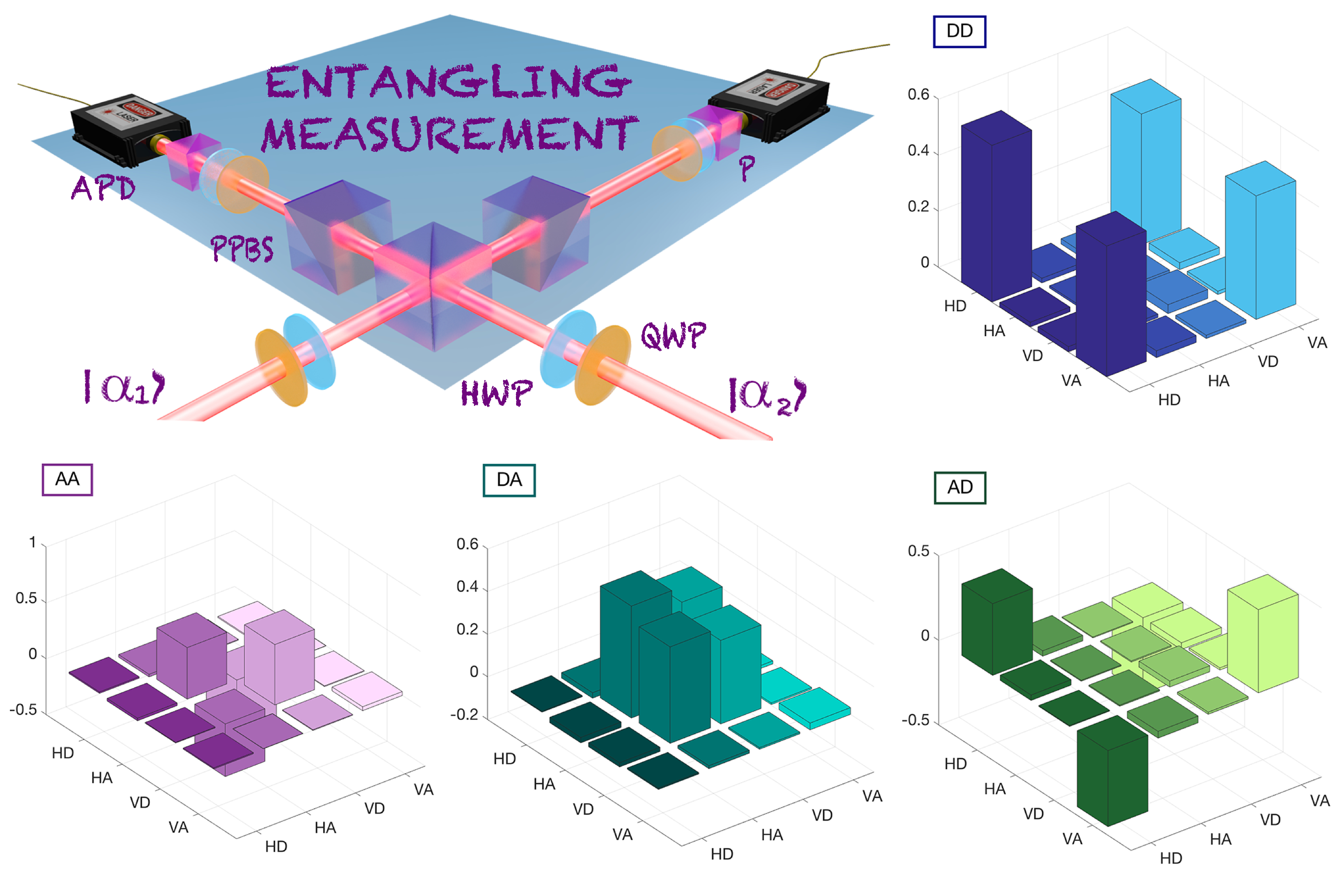}
\caption{Implementation and characterisation of the entangling measurement. The entangling operation is realised by means of a partially-polarising beam splitter (PPBS): the additional phase on the $\ket{V}_1\ket{V}_2$ component results from polarisation-dependent quantum interference~\cite{Langford05,Kiesel05,Okamoto05}. The photons are both measured along the polarisations $D$ and $A$; each combination corresponds to a distinct Bell state. This is accomplished by means of a half-wave plate (HWP), a polariser (P), and an avalanche photodiode (APD), with an additional a quarter-wave plate (QWP) needed to compensate for birefringent phase shift from the PPBS. In order to characterise the response of the measurement device, we input a set of reference states $\ket{\alpha_1}$ and $\ket{\alpha_2}$. By inverting Born's rule for the detection probability, one gets the expressions for the actual matrices $\Pi_{DD}$ (fidelity $F=(96.72\pm0.19)\%$ with the ideal case), $\Pi_{AA}$ ($F=(97.26\pm0.13)\%$), $\Pi_{DA}$ ($F=(91.11\pm0.29)\%$), and $\Pi_{AD}$ ($F=(91.01\pm0.34)\%$). The Methods section contains details on the experimental procedures.}
\label{Figura2}
\end{figure*}

Here we consider the entangling measurement shown in \ref{Figura2}, based on the use of a linear-optical Control-Sign (CS) gate~\cite{Langford05,Kiesel05,Okamoto05} acting on two polarisation qubits, each of the form $\alpha\ket{H}+\beta\ket{V}$ ($H$ and $V$ are the horizontal and vertical polarisations, respectively). The CS gate imparts a $\pi$-phase shift to the $\ket{V}_1\ket{V}_2$ with respect to the other three combinations $\ket{H}_1\ket{V}_2$, $\ket{V}_1\ket{H}_2$, and $\ket{H}_1\ket{H}_2$. If we consider its action in the rotated basis $\ket{D}=1/\sqrt{2}(\ket{H}+\ket{V})$, $\ket{A}=1\sqrt{2}(\ket{H}-\ket{V})$, the maximally entangled (Bell) state $1/\sqrt{2}\left(\ket{H}_1\ket{D}_2+\ket{V}_1\ket{A}_2\right)$ is transformed into the separable state $\ket{D}_1\ket{D}_2$; similarly, the other three states $1/\sqrt{2}\left(\ket{H}_1\ket{D}_2-\ket{V}_1\ket{A}_2\right)$, $1/\sqrt{2}\left(\ket{H}_1\ket{A}_2+\ket{V}_1\ket{D}_2\right)$, and $1/\sqrt{2}\left(\ket{H}_1\ket{A}_2-\ket{V}_1\ket{D}_2\right)$ are mapped onto separable states forming an orthogonal set (see Methods). Therefore, polarisation analysis performed after the gate is equivalent to a discrimination of these four possible Bell states. We have implemented one such device, and tested its capabilities by means of detector tomography~\cite{Lundeen09}: one constructs a matrix $\Pi_{D,D}$, such that the detection of a $DD$ event occurs with a probability $p(D,D)=\tr{\Pi_{D,D}\rho}$ for any input state $\rho$, and likewise for the other three instances. In the ideal limit, the detection matrices correspond to projectors onto Bell states. Fig.~\ref{Figura2} shows the four experimental matrices, which resemble closely the expected states (see Methods).

The knowledge of the matrices allows us to evaluate the Fisher information for a generic estimation problem. Here we will detail two cases, phase with diffusion and multiple phases, showing how they are in fact intrinsically different, and how this difference is manifested when one considers entangling measurements. While one can show that all the results we will present here hold for generic qubit states, for the sake of simplicity we will focus on initial equatorial states in the $yx$-plane, i.e. $|\psi_0\rangle = ( |0\rangle + e^{i \xi}|1\rangle)/\sqrt{2}$; due to the functioning of the gate, $\{|0\rangle, |1\rangle\}$ correspond to $\{|H\rangle_1, |V\rangle_1\}$ for one of the qubits, and to $\{|D\rangle_2, |A\rangle_2\}$ for the second in our polarisation coding.

As for the unitary part, we consider rotations in the form $ R(\phi_y,\phi_z) = \exp \{ i ( \phi_y \sigma_y + \phi_z \sigma_z)\} $ applied to the initial state $|\psi_0\rangle$. When this undergoes a single phase rotation along $z$-axis $R(0,\phi$) and is also subjected to a dephasing evolution characterized by the parameter $\delta$ \cite{Genoni12,Vidrighin14}, the output mixed state reads
$$
\rho_{\phi,\delta} =  \frac12 \left(
\begin{array}{c c}
1 & e^{-i (\phi+\xi) - \delta^2} \\
e^{ i (\phi+\xi) - \delta^2}  & 1
\end{array}
\right).
$$

\begin{figure}[t!]
\includegraphics[width=\columnwidth]{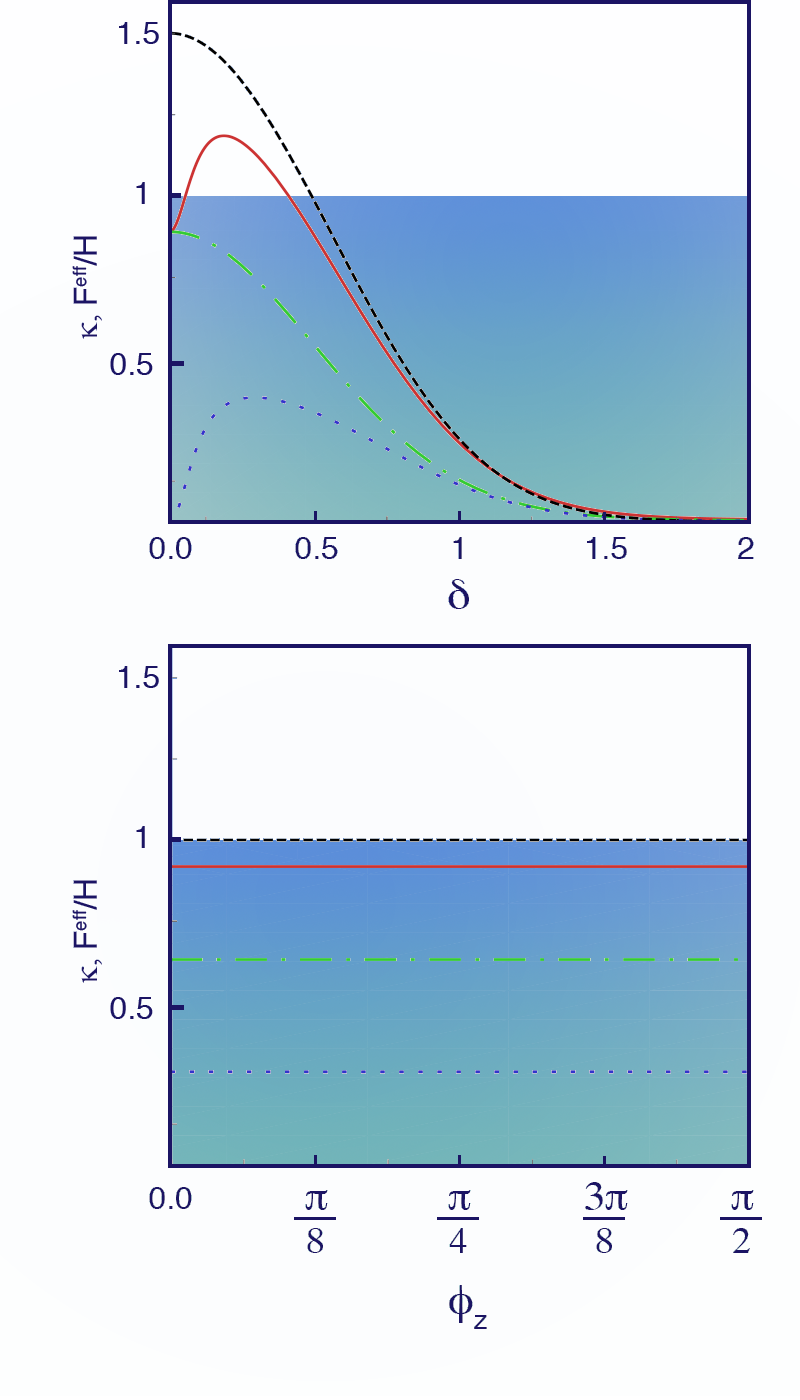}
\vskip -9mm
\caption{Fisher information from the experimentally reconstructed detector matrices. Upper panel: phase and phase diffusion estimation with our entangling measurement. The red solid curve is the experimental $\kappa$ for the value $\phi\simeq0.89$ delivering the highest FI, optimised over the inputs. A numerical search outputs: $\xi=0$ for the first qubit, and $\xi=0.10$ for the second. This is composed of the two contributions of the effective FI for the phase (dash-dotted green line), and for the dephasing (dotted blue line), normalised to the corresponding QFI. The black dashed curve is a prediction for the ideal case, and the shadowed area $\kappa\leq1$ is accessible with single-qubit measurements. The optimal value is $\kappa^*=1.18\pm0.02$, nine standard deviations above the limit. Lower panel: two-phase estimation with the same entangling measurement. The colour code is as above, except that the dash-dotted green line corresponds to $\phi_y=0$, and the blue dashed curve is associated to $\phi_z$. In this case, the initial phases $\xi_1$ and $\xi_2$ of the input probes are identical, and optimised for the estimation of $\phi_z$.}
\label{Figura3}
\end{figure}

We report the value of $\kappa$ in Eq.\eqref{kappa} estimated using our measurement device in Fig.~\ref{Figura3}; this is shown as a function of $\delta$ for the value of $\phi$ delivering the best performance, along with the prediction for the ideal case. There exist a range of values for $\delta$ for which an improvement over the independent strategy ($\kappa\leq1$) is assessed; however, in the region where the largest amelioration is expected ($\delta\simeq0$), we actually assist to a drop in the information. In fact, the ideal performance in this range is enhanced by the highly symmetric repartition of the counts among the four possible outcomes; once this condition is spoilt by the experimental imperfections, the improvement is compromised beyond any repair, and, in particular the ability of the detector of estimating $\delta$ goes to zero. One can mitigate the impact of non-idealities by biassing the input states so to achieve best performance, however this is effective in the low $\delta$ regime.  

The other difference with respect to the independent strategy concerns the applicability of entangling measurements. These need a good level of coherence to deliver a meaningful estimation, while the separable case works independently on the value of $\delta$. This is manifestation of the fact that collective strategies are more sensitive to quantum signatures of the states when delivering an advantage~\cite{Gu12,Girolami14}: while these previous studies have highlighted the role of correlations, our work indicates that this even occurs at the single-qubit level. We notice that one can not attack this with the usual adaptive approach, since we are in the presence of dissipation and this can not be recovered.

We now turn our attention to the purely unitary case, where we seek to estimate the parameters $\phi_y$ and $\phi_z$; no enhancement is observed in this case. Therefore, we can conclude that, while there exist a universal bound applying to the single qubit case, this can not be addressed by means of entangling measurement in any instance; further inspection is needed to assess whether this might be the case.

\section{Discussion}  
The usefulness of collective strategies is simply captured by the weak commutativity condition Eq.~\eqref{eq:cond}. The two SLD operators corresponding to the parameters $\phi_y$ and $\phi_z$ can be easily calculated and one obtains that the condition (\ref{eq:cond}) is never satisfied, unless for a specific initial phase $\xi= \bar{\xi}(\phi_y,\phi_z)$. However, by picking this initial phase, one observes that the output state $|\psi_{\phi_y,\phi_z}\rangle = R(\phi_y,\phi_z)|\psi_0\rangle$ still corresponds to an equatorial state, like the initial state $|\psi_0\rangle$, where all the information on the two parameters is cast on a new relative phase $\xi^\prime$. As expected, the corresponding QFI matrix is singular, as it is not possible to estimate two parameters from only the relative phase characterising a pure qubit state. In general, we have thus found that it is impossible to saturate the QCR bound $\kappa \leq 2$ for a two-phase estimation problem, no matter how many copies of the qubit probe state are accessed with entangling measurements. On the other hand, the condition (\ref{eq:cond}) involving the SLD operators for the parameters $\phi$ and $\delta$ is always satisfied with a non-singular QFI matrix $H$. 

These results show on the one side how the two estimation problems are intrinsically different; however, on the other hand they do not tell us anything about the possible intermediate values of the figure of merit $\kappa$ that might be achievable when one considers measurement strategies involving multiple copies of probe states. In the phase-dephasing estimation case, considering collective measurements on two copies at the time, one obtains an ameliorated bound $\kappa\leq1.5$~\cite{Vidrighin14}, achieved by exploiting a Bell measurement. Performing the same numerical investigation for the two-phase estimation problem, one finds that the {\em one-copy} bound $\kappa \leq 1$ seems to be satisfied regardless how much entangling are the measurements we consider on two copies of the quantum state. Besides showing again the paradigmatic difference between the two estimation problems, this observation leads us to the conjecture that the bound $\kappa\leq 1$ is satisfied for the two-phase estimation for any number of copies $m$ of the probe state that we decide to measure jointly.

Some considerations are {\it \`a propos} for our photonic implementation. In this proof-of-principle, we have neglected the fact that the gate only succeeds a fraction of the times, since it relies on post-selection, albeit this might be mitigated by pre-biassing the inputs~\cite{Langford05,Kiesel05,Okamoto05}; future application in optical sensors of our technique will have to rely on progress in quantum photonics devices, however this might find immediate application in quantum-enhanced sensors for magnetometry~\cite{Meyer01,Thiel16,Choi17}. 


In conclusion, we have evaluated the usefulness of entangling measurements to implement joint estimation with two qubit probes. We have shown that the case of phase and phase diffusion can be improved over the case of independent measurements, under opportune conditions. This however is not a general result as shown by the counterexample of multiple phases. This asymmetry is not revealed in the single qubit regime, since both cases obey the same bound. We traced the origin of these disparate behaviours in the weak commutativity condition, which inspects the SLD operators, therefore, the infinitesimal generators of the transformation. Our results reinforce the view that SLDs calculated in the single-probe case encompasses most information on the applicability of collective measurements.

\section{Methods}

{\bf Experimental details.} We use a photon pair source based on parametric downconversion; this consists of a 2-mm barium borate crystal, pumped with a 405nm laser diode; frequency-degenerate photons emitted with a 5$^{\circ}$ angle are coupled into single-mode fibres, and delivered to the CS gate. Spectral filtering is applied with Gaussian filters with 7.5-nm width (full width half maximum). Observed coincidence rates are of the order of 1500 coincidences/s. The working principle of the CS gate is polarisation-dependent two-photon interference: the PPBSs in Fig.~\ref{Figura2} have unequal transmittivity $t_H=1$ and $t_V=1/\sqrt{3}$ for the two polarisation directions. The probability amplitude that two photons emerge on different output arms is: $t_{x1}t_{x2}-r_{x1}r_{x2}$, where $x1=H,V$ and $x2=H,V$; the choice of transmittivities result in the appearance of the extra $\pi$ shift, conditioned on post-selecting a coincidence detection. In order to compensate for the unequal amplitudes, we inserted one more PPBS on each arm rotated by 90$^{\circ}$, so that the role of horizontal and vertical polarisations are exchanged~\cite{Pryde}: the output probability then becomes polarisation insensitive.

{\bf Detector tomography}. The characterisation of the detector is carried out by associating to a matrix $\Pi_k$ such that for any two-qubit input state $\rho^{(2)}$ the probability of observing the outcome $k$ is $p(k)=\text{Tr}\left[\rho^{(2)}\Pi_k\right]$. These matrices must be non-negative $\Pi_k \geq 0$ and they must sum to the identity operator $\sum_k \Pi_k={\mathbf I}$. The reconstruction algorithm takes as the input the experimental probabilities 
for products of single-qubit reference states $\ket{\alpha_1}\ket{\alpha_2}$ chosen among $\{\ket{H},\ket{V},\ket{D},\ket{A},\ket{R},\ket{L}\}$, ($R$ and $L$ are the two circular polarisations). The algorithm then proceeds to find the closest set $\{\Pi_k\}$ that fits the data to the expected values $\text{Tr}\left[\ket{\alpha_1}\bra{\alpha_1}\otimes\ket{\alpha_2}\bra{\alpha_2}\Pi_k\right]$ with a maximum likelihood routine, constrained to the physical requirements on the matrices. Uncertainties on evaluated quantities are calculated by means of a Monte Carlo routine that simulates multiple runs of the reconstruction experiment by varying at each run the detected coincidence counts within its uncertainties.

\vskip 1mm
{\bf Acknowledgements.} The authors thank M.A. Ricci and B. Smith for valuable discussions, and G. Pryde for consultancy on the setup. This work has been supported by the EC project QCUMbER (grant no. 665148). MGG acknowledges support from Marie Sk\l odowska-Curie Action H2020-MSCA-IF-2015 (project ConAQuMe, grant nr. 701154). {\bf Author contributions} MB and MGG conceived the project. ER and LM designed the setup, and performed the experiment with IG and MS. ER and MGG carried out the data analysis with inputs from IG, MS, FS, and MB. All the authors discussed the results and their interpretation and wrote the article. {\bf Competing financial interests.} The authors declare no competing financial interests.

\end{document}